\documentclass[%
 reprint,
superscriptaddress,
 amsmath,amssymb,
 aps,
pra,
floatfix,
]{revtex4-2}

\usepackage{graphicx}% Include figure files
\usepackage{dcolumn}% Align table columns on decimal point
\usepackage{bm}% bold math
\usepackage{textcomp}
\usepackage{gensymb}

\usepackage{amsmath}
\usepackage{booktabs}  % For better table formatting
\usepackage{multirow}
\usepackage{makecell}  % For multi-line cells

\usepackage{xcolor}
\usepackage{hyperref}

\begin{document}

\preprint{APS/123-QED}

\title{Daylight quantum keyless private communication for free-space links}

\author{Pedro Neto Mendes}
\affiliation{Quantum Physics of Information Group, Instituto de Telecomunica\c{c}\~oes, Av. Rovisco Pais, 1049-001 Lisbon, Portugal}
\affiliation{Quantum Information and Quantum Optics Laboratory, Instituto Superior T\'ecnico, 1049-001 Lisbon, Portugal}

\author{Preeti Yadav}
\affiliation{Quantum Physics of Information Group, Instituto de Telecomunica\c{c}\~oes, Av. Rovisco Pais, 1049-001 Lisbon, Portugal}

\author{Lourenço Sumares}
\affiliation{Quantum Information and Quantum Optics Laboratory, Instituto Superior T\'ecnico, 1049-001 Lisbon, Portugal}

\author{Hugo Zbinden}
\affiliation{Vigo Quantum Communication Center, University of Vigo, Vigo E-36310, Spain}
\affiliation{Escuela de Ingeniería de Telecomunicación, Department of Signal Theory and Communications, University of Vigo, Vigo E-36310, Spain}
\affiliation{AtlanTTic Research Center, University of Vigo, Vigo E-36310, Spain}

\author{Davide Rusca}
\affiliation{Vigo Quantum Communication Center, University of Vigo, Vigo E-36310, Spain}
\affiliation{Escuela de Ingeniería de Telecomunicación, Department of Signal Theory and Communications, University of Vigo, Vigo E-36310, Spain}
\affiliation{AtlanTTic Research Center, University of Vigo, Vigo E-36310, Spain}

\author{Emmanuel Zambrini Cruzeiro}
\email{emmanuel.cruzeiro@lx.it.pt}
\affiliation{Quantum Physics of Information Group, Instituto de Telecomunica\c{c}\~oes, Av. Rovisco Pais, 1049-001 Lisbon, Portugal}
\affiliation{Quantum Information and Quantum Optics Laboratory, Instituto Superior T\'ecnico, 1049-001 Lisbon, Portugal}

\date{\today}

% ---------- Abstract ----------
\begin{abstract}
Quantum key distribution (QKD) is the most established approach in quantum communication. However, long-distance free-space implementations, particularly satellite links, remain challenging, especially during the day due to daylight background noise. Quantum keyless private communication (QKPC) is a quantum communication protocol that enables information-theoretic security with simpler system requirements, improved robustness against noise, and without the need for secret key distribution. QKPC and QKD are complementary, with QKPC enabling free-space links where QKD is impractical, while QKD provides channel monitoring for applications that require eavesdropping detection. Here, we report a complete implementation of QKPC in a daylight free-space experiment over a 90 m rooftop link, using an experimentally simple setup. Our demonstration includes all stages of the protocol, from encoding and synchronization to message decoding, operates entirely without auxiliary classical synchronization channels, and is implemented offline through post-processing. This work demonstrates the feasibility of practical and scalable quantum communication over high-noise daylight links and highlights the potential of QKPC as a complementary solution to QKD for future ground-based and space-based communication systems.
\end{abstract}

%\begin{abstract}
%Quantum key distribution (QKD) is the most established approach in quantum communication, but long-distance free-space implementations, particularly satellite links, remain challenging under daylight background noise. Quantum keyless private communication (QKPC) offers a complementary approach, enabling information-theoretic security without secret key distribution, with simpler system requirements and improved noise robustness. Here, we report a complete daylight free-space implementation of QKPC over a 90 m rooftop link. The experiment demonstrates encoding, synchronization, and message decoding using an experimentally simple setup, without auxiliary classical synchronization channels, with synchronization performed offline through post-processing. These results demonstrate the feasibility of QKPC for practical free-space quantum communication.
%\end{abstract}

\maketitle

% =====================================================
\section{Introduction}
% =====================================================

%\cite{google}
Quantum key distribution (QKD) is the most widely studied and deployed quantum communication protocol, enabling two distant parties to generate shared secret keys that can subsequently be used for secure communication. Over the past decades, QKD has matured into a commercially available technology and is being deployed in national and international quantum communication networks \cite{QKD_review}. Nevertheless, QKD faces important practical limitations in free-space scenarios: daylight operation remains challenging \cite{Nature_daylight_long_range_QKD, Nature_daylight_free_QKD, Optica_free_space_QKD}, and satellite-based implementations typically achieve low secure key rates while requiring complex and resource-intensive systems \cite{Optica_simulation_improve_qkd_satellite, Nature_QKD_satellite_analysis, science_QKD_satellite_analysis}.

Quantum keyless private communication (QKPC) has been proposed as a complementary quantum communication protocol that exploits physical-layer asymmetries in the eavesdropper channel, enabling direct secure transmission of messages \cite{Original_QKPC_paper}. Although its security model is less general than that of QKD, QKPC offers several practical advantages, including simpler system requirements, the elimination of key management and storage, as well as the classical encryption/decryption stage post key distribution, and an increased tolerance to background noise \cite{Ploarization_QKPC}. These features make QKPC particularly useful for daylight free-space and satellite communication scenarios. We envision future satellite systems capable of supporting both QKD and QKPC for a near-term deployment of long-distance quantum communication infrastructures \cite{Cubesat_QKPC}.

QKPC can be implemented using different optical degrees of freedom. In the simplest case of on-off keying (OOK), information is encoded in the presence or absence of photons, which makes the protocol infeasible when background noise becomes comparable to the signal. Polarization-based QKPC requires a more complex setup but allows communication in noise regimes where OOK would not be secure.

In this work, we experimentally demonstrate a complete implementation of the polarization QKPC protocol over a rooftop free-space link (90 meters) operating under daylight conditions. The system includes classical encoding and decoding, and operates without auxiliary synchronization channels. The paper is organized as follows: Section \ref{sec:SectionII} describes the implemented protocol and system architecture, Section \ref{sec:SectionIII} presents the experimental setup, Section \ref{sec:SectionIV} reports the results, and Section \ref{sec:SectionV} concludes the paper.

% =====================================================
\section{Protocol Implementation}
\label{sec:SectionII}
% =====================================================

\subsection{Quantum Keyless Private Communication}

QKPC, introduced in \cite{Original_QKPC_paper}, enables the direct secure transmission of information without relying on pre-shared secret keys. Security is based on the classical–quantum wiretap channel model \cite{wyner, csiszar}, where confidentiality arises from a physical-layer asymmetry between the legitimate receiver, Bob, and a potential eavesdropper, Eve.
%\textcolor{blue}{Unlike some direct communication protocols, such as quantum secure direct communication (QSDC) \cite{pan2024evolution}, QKPC does not involve two-way or two-step quantum transmission. These QSDC protocols, therefore, suffer from additional channel losses, making their transmission rates even lower than QKD. On the other hand, QKPC can achieve even higher transmission rates than QKD \cite{Original_QKPC_paper}. While QKPC enables single-shot communication capability, it lacks explicit eavesdropper detection of QSDC protocols. Moreover, QKPC does not rely on interactive classical communication during and post the quantum transmission stages, unlike QKD and QSDC. QKPC merely requires interactive communication during the initial setup phase (e.g., channel characterization and parameter agreement), as is the case with most quantum/classical communication protocols.}
Unlike quantum secure direct communication (QSDC), where two-way or two-step quantum transmission results in higher channel losses, QKPC operates in a one-shot configuration, enabling higher transmission rates than both QKD and QSDC. However, QKPC
does not include explicit eavesdropper detection, trading
active monitoring for one-shot communication capability \cite{Original_QKPC_paper}. Moreover, QKPC avoids interactive classical communication during and after the quantum transmission stage, unlike both QKD and QSDC, requiring only minimal interaction during the initial setup phase for channel characterization and parameter agreement.
%Unlike QKD, which relies on interactive classical communication to establish a secret key, QKPC operates in a one-way configuration, requiring no classical communication during or after the quantum transmission stage, and only interactive communication during the initial setup phase (e.g., channel characterization and parameter agreement).
%Moreover, unlike some quantum secure direct communication (QSDC) protocols (another class of quantum protocols for direct communication), QKPC does not include explicit eavesdropper detection, trading active monitoring for one-shot communication capability.

In the wiretap framework, two legitimate distant users, Alice and Bob, communicate over a physical channel, while Eve, the wiretapper, has access to a degraded version of the same signal. The channel performance is quantified by the private capacity, $C_P$, defined as the maximum rate at which information can be reliably decoded by Bob while remaining inaccessible to Eve. The asymmetry between Bob and Eve may originate from differences in channel loss, noise, or interception capability. In free-space links, this naturally results from geometric constraints, finite receiver apertures, and limited collection efficiency.

In this work, we adopt the same security framework used in previous QKPC demonstrations \cite{Original_QKPC_paper, QKPC_Decoy, Ploarization_QKPC}. The goal of the present experiment is not to analyze security against arbitrary interception strategies or to optimize the performance in terms of the communication rate, but to validate a complete implementation under realistic daylight conditions, including synchronization, decoding, and classical pre and post-processing.

The communication link is modeled as a single-mode free-space bosonic channel with overall transmission efficiency $\eta$, incorporating propagation loss, turbulence, pointing errors, and receiver inefficiencies. Eve’s channel is assumed to be degraded by a factor $\gamma \in (0,1)$, such that her effective transmission efficiency is $\gamma \eta \le \eta$, where $\eta$ is the channel transmittance. A detailed description of this channel model can be found in \cite{Original_QKPC_paper}.

For security analysis, Eve is assumed to perform optimal quantum measurements achieving the Helstrom bound, representing a worst-case discrimination strategy. Bob, in contrast, employs threshold single-photon detectors with finite efficiency and nonzero dark count probability.

\subsection{Implementation}
\label{sec:IIB}

\begin{figure}[h!]
  \centering
  \includegraphics[width=0.48\textwidth]{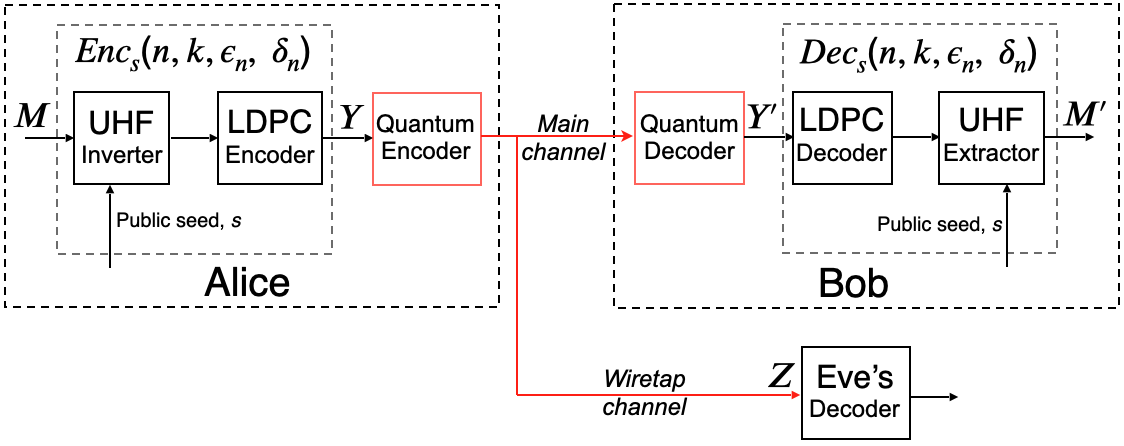}
  \caption{The wiretap coding scheme. Alice and Bob communicate over a noisy main channel, whereas Eve's wiretap channel is a degraded channel. Bob receives a noisy version $Y'$ of the transmitted codeword $Y$, whereas Eve receives $Z$ with a degradation factor $\gamma$.}
  \label{fig:encoder}
\end{figure}

In this work, we provide a full implementation of QKPC. The classical pre-processing involves mapping a binary message $M$ of length $k$ to a longer sequence $Y$ of length $n$ using a publicly known wiretap encoder $Enc_s(n,k,\epsilon_n,\delta_n)$, where $\epsilon_n$ is the error probability and $\delta_n$ is the information leakage, as shown in Figure \ref{fig:encoder}. 
The wiretap coding scheme allows secure and reliable transmission of a message over a noisy channel \cite{wyner}. The coding scheme used consists of two parts: secure encoder (decoder) and error-correction encoder (decoder). Universal hash family (UHF) based secure coding followed by an error-correction code results in mapping the message $M$ of length $k$ to a codeword $Y$ of length $n$. To ensure reliability and secrecy of the transmitted message $M$, the wiretap coding rate $R=k/n$ is chosen such that $k/n < C_P$. Upon receiving a noisy version of transmitted codewords $Y'$, the receiver applies the error-correction decoder followed by UHF based secure decoder to obtain the message $M'$, such that $M'=M$ with probability $1-\epsilon_n$. We use an information-theoretically secure and computationally efficient construction of UHF based invertible extractor using a random Toeplitz matrix with a public seed $s$ \cite{hayashi2011exponential, UHF}. For error-correction, a low-density parity check (LDPC) code is used where the matrices were generated using the python library Pyldpc \cite{pyLDPC}, while the decoding using the library ldpc \cite{bpldpc}.

For the transmission of the codewords, we employ the polarization-based QKPC scheme introduced in \cite{Ploarization_QKPC}. This choice increases the complexity of the experimental setup, but enables communication in regimes with stronger background noise, including daylight conditions. The logical bits are encoded in weak coherent states with different linear polarizations,
\begin{equation}
\begin{aligned}
|\psi_0 \rangle &= |\alpha\rangle_H \otimes |0\rangle_V, \\
|\psi_1 \rangle &= |\alpha \cos(\theta)\rangle_H \otimes |\alpha \sin(\theta)\rangle_V,
\end{aligned}
\label{Eq: Polarization states simplified}
\end{equation}
where subscripts $0$ and $1$ label the logical bits, $H$ and $V$ represent the horizontal and vertical polarization modes, $\alpha$ is the coherent-state amplitude, and $\theta$ determines the polarization separation between the two states.

Both states are prepared with the same total mean photon number, $|\alpha|^2_0=|\alpha|^2_1=|\alpha|^2$, but with different photon-number distributions between the two polarization modes. In particular, $|\psi_0\rangle$ is purely horizontally polarized, while $|\psi_1\rangle$ is rotated by an angle $\theta$ in the polarization plane. This constant photon-number constraint ensures that the encoding relies mostly on polarization and not on intensity modulation.

At the receiver, a polarization rotation, $\Omega=-\frac{\theta}{2}+\frac{\pi}{4}$, is applied to minimize the discrimination error probability. The measurement consists of independently detecting the photon numbers in the horizontal and vertical modes and performing a minimum-error decision based on the relative counts, estimating which state was most likely transmitted.

To account for background light and detector noise, we model an independent Poissonian noise contribution with mean photon number $\Delta$ per detector per bit. This additional noise directly impacts the discrimination error probability and the resulting quantum bit error rate (QBER).

Since the emitter and receiver operate with independent clocks and no shared timing reference signal, synchronization is performed directly from the quantum signal itself, similar to other works \cite{Qbit_4_sync, krause2025clock}. We concatenate known sequences of bits, called preamble henceforth, before each transmitted codeword of a given length. For the sake of simplicity, we choose the same preamble and codeword sequence to be transmitted repeatedly. The preamble was constructed to avoid internal periodic or repeated patterns, ensuring a single dominant correlation peak and preventing ambiguous temporal alignment.

The goal of synchronization is for the receiver to, first, correct the relative frequency drift between their clock and the transmitter's by recovering the period. Secondly, upon finding the transmitter's clock period, to determine the time-offset.
The nominal bit period $T_b$ is obtained from the measured bit rate via a Fourier analysis of a subset of the data. For a trial temporal offset $\delta$, the time axis is divided into intervals of duration $T_b$, and photon counts are accumulated per interval, $n_H(i)$, $n_V(i)$,
where $i$ denotes the bit index. A preliminary decoded sequence $\hat{b}(i)$ is obtained by assigning each bit according to the detector with the larger photon count.

Let $b_{\mathrm{pre}}(j)$ denote the known preamble of length $L_{\mathrm{pre}}$. For each candidate offset $\delta$, the correlation,
\begin{equation}
C(\delta) = \sum_{j=1}^{L_{\mathrm{pre}}}
\left( 1 - 2 \left| \hat{b}_{\delta}(j) - b_{\mathrm{pre}}(j) \right| \right),
\end{equation}
is computed. The value of $\delta$ that maximizes $C(\delta)$ defines the optimal alignment. We do not optimize over how frequently the preambles must be inserted in the transmitted data, nor the ratio of the preamble length and the codeword length. %Moreover, the smallest value of $L_{\mathrm{pre}}$ necessary depends on $\eta$, and must be such that $L_{\mathrm{pre}} >>1/\eta$ \cite{Qbit_4_sync}. Since QKD works at a lower average photon number than QKPC, this brings an advantage in the minimum length of the preamble.

To account for residual clock mismatch, a fine scan of $T_b$ around its nominal value is performed, and the pair $(\delta, T_b)$ that maximizes the preamble correlation is selected. Once the preamble peak is identified, the start of the codeword is inferred from the known preamble length, and the transmitted bits are extracted using the synchronized boundaries.

% =====================================================
\section{Experimental Implementation}
\label{sec:SectionIII}
% =====================================================

\begin{figure}[t]
\centering
\includegraphics[width=0.8\linewidth]{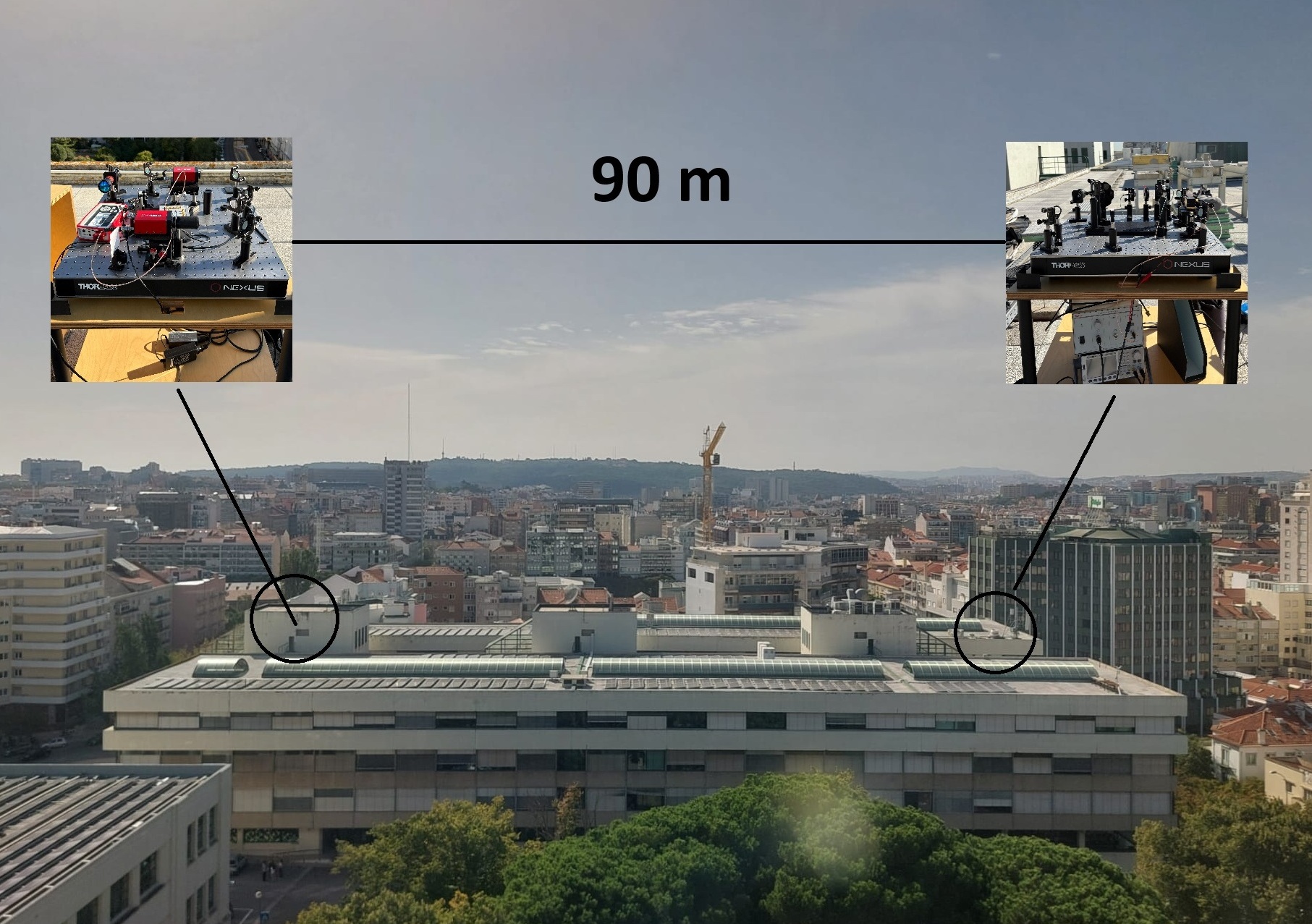}
\caption{
Photograph of the 90 m rooftop free-space link used in the experiment.
The transmitter (right) and receiver (left) were on the same rooftop,
operating under daylight urban conditions. Insets show the optical setups used at each position.
}
\label{fig:rooftop_link}
\end{figure}

The experimental setup used in this work is based on the polarization QKPC setup previously reported in~\cite{Ploarization_QKPC}, with minor modifications to enable portable outdoor operation. All optical components of both the transmitter and receiver are mounted on compact breadboards placed on mobile tables, allowing the system to be transported and deployed on the rooftop of the Instituto Superior Técnico university building in Lisbon, Portugal, Figure~\ref{fig:rooftop_link}. The experiment was performed at increasing transmitter–receiver separations, starting from shorter distances and extending up to the maximum available rooftop distance of 90 m.

At the transmitter, optical pulses with a beam diameter of approximately 5 mm are generated using a gain-switched laser at 850 nm (Roithner VC850MZ), driven by a function generator (AIM-TTI TG330) at a repetition rate of 50 kHz. The pulses, with a duration of 10 $\mu$s, corresponding to a 50\% duty cycle are directed through a polarizing beam splitter (PBS) (Thorlabs PBS102), where one polarization component is monitored using a power meter (Thorlabs S120VC) to characterize the emitted optical power. The horizontal polarization component is sent through an electro-optic polarization modulator (EOPM, Thorlabs EO-AM-NR-C1), driven by a high-voltage amplifier (Thorlabs HVA200), and followed by a polarization controller composed of two quarter-wave plates (Thorlabs WPQ05M-850) and one half-wave plate (Thorlabs WPH05M-850). This configuration enables the preparation of the well-defined linear polarization states on a pulse-by-pulse basis.

\begin{figure}[ht]
  \centering
  \includegraphics[width=0.48\textwidth]{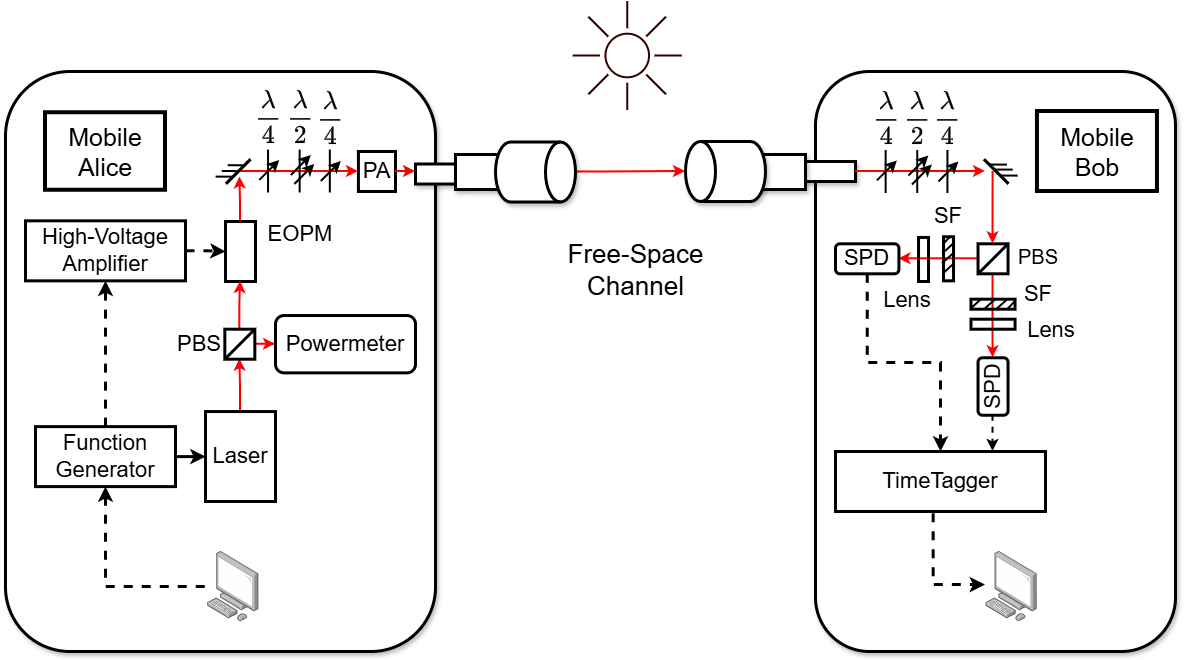} 
  \caption{Mobile experimental setup for the full implementation of polarization QKPC. PBS: polarizing beam splitter. EOPM: electro-optic polarization modulator. PA: passive attenuator. SPD: single photon detector. $\frac{\lambda}{4}$: quarter-waveplate. $\frac{\lambda}{2}$: half-waveplate. SF: spectral filter. The black dashed arrows are electrical connections while the red solid arrows represent the optical signal.}
  \label{fig:setup}
\end{figure}

The optical pulses are subsequently attenuated using a combination of neutral density filters (Thorlabs NE510B-B) to produce weak coherent states with a controlled average photon number. A schematic of the experimental setup is shown in Figure~\ref{fig:setup}.

Both the transmitter and receiver are equipped with $3\times$ beam expanders (Thorlabs GBE03-B) to reduce the beam divergence over the free-space link. After expansion, the beam diameter along the link is around 20 mm. Since the propagation distance is relatively short, the additional increase in beam size due to diffraction is minimal. At the receiver, the collected beam passes through a polarization controller, which compensates for channel-induced polarization rotations and aligns the received states with the measurement basis that maximizes the discrimination probability.

At the receiver, the incoming light is then separated using a PBS and detected by two threshold single-photon detectors (Thorlabs SPDMH3). Each detector has an active-area diameter of 100 $\mu$m, a photon detection efficiency of approximately 50\%, a dark count rate of $\sim 70$ Hz, and a dead time of $\sim 45$ ns. To improve the receiver efficiency, 75 mm focal-length lenses are used to focus the beam onto the detector active areas. Taking into account the 100 $\mu$m detector diameter, the 75 mm focusing lens, and the receiver $3\times$ beam expander used in reverse, the full-angle field of view of each detection channel is approximately $0.44$ mrad. At the maximum link distance of 90 m, this corresponds to a transverse field-of-view diameter of about 4 cm. Detection events are recorded using a time-tagging unit (Swabian Time Tagger 20), which registers the arrival time and detector channel of each click. To suppress background light, a spectral filter centered at 850 nm with a bandwidth of 10 nm (Thorlabs FBH05850-10) is placed in front of the detectors within a lens tube.

The average number of photons arriving at the receiver is initially estimated from the measured optical power at the transmitter, accounting for channel and receiver losses. After recording the data, a precise estimation is made using the photon statistics recorded.

All experimental data is processed offline. After synchronization, comparing the recovered sequence with the expected codeword allows estimation of the QBER, which, together with the measured photon statistics, is used to infer the private capacity under the adopted wiretap model.

Although the detectors used in this experiment are threshold single-photon detectors, the long pulse duration allows us to approximate a photon-number-resolving measurement in post-processing, as previously demonstrated in~\cite{Ploarization_QKPC}. For each transmitted bit, the time tagger records all detection events occurring within the corresponding temporal window, and we count the clicks registered by each detector to obtain the effective photon counts. Since the pulse duration is much longer than the detector dead time, several independent clicks can be registered within one bit interval. The bit value is then estimated by comparing the two accumulated counts. While well suited for characterization and proof-of-principle demonstrations, increasing the pulse width enlarges the detection window and consequently increases background noise.

% =====================================================
\section{Results and discussion}
\label{sec:SectionIV}
% =====================================================

The dataset analyzed corresponds to an experimental acquisition performed in Lisbon (IST Alameda) on 23 September 2025 between 17:00 and 17:30 local time, Figure \ref{fig:rooftop_link}. Each transmission consisted of a 10000 bit preamble followed by a 50000 bit codeword (60000 bits total). The codeword length, $n$, was chosen in a conservative way, so as to allow enough redundancy in the original message length $k$ to allow for a reliable error-correction, while keeping the message secret. These blocks were repeated at a 50 kHz bit rate to make 10 acquisitions, where each set was 10 seconds long, with approximately 2 minutes of idle interval between consecutive sets. In total, 71 complete blocks were transmitted and detected, corresponding to 4.26 Mbits of exchanged data at a 50 kHz bit rate.

For the experiment, we used approximately 15 detected photons per pulse and a polarization separation of $60^\circ$. These parameters were chosen because they are close to the expected optimal regime for the rooftop noise conditions, while remaining straightforward to implement experimentally. The channel attenuation was estimated to be 7 dB, with an additional receiver loss of 10 dB including coupling and detector efficiency, based on calibration measurements from similar runs.

The experiment was conducted under daylight conditions with mostly clear skies (25$^\circ$C, 19\% relative humidity, 3.4 m/s wind). The line of sight of both detectors passed close to the sun, resulting in high background levels. Background count rates were estimated during the idle intervals between blocks.

To evaluate the synchronization performance, we compare the following two cases: (A) 50000 bit codeword-based alignment where the entire codeword was used; (B) 500 bit preamble-based alignment (as discussed in Section \ref{sec:IIB}). The codeword-based alignment is used only as a benchmark, while the 500 bit preamble was chosen as the shortest preamble length that allowed reliable synchronization without losing detected blocks.
%(A) 50000 bit codeword-based alignment; (B) 500 bit preamble-based alignment. The codeword-based alignment, where the entire codeword was used for synchronization, serves only as a benchmark. Whereas the 500 bit preamble-based alignment (as discussed in Section \ref{sec:IIB}) is the implemented synchronization. The 500 bit preamble was chosen as the shortest preamble length that allowed reliable synchronization without losing detected blocks.

\begin{figure}[ht]
\centering
\includegraphics[width=0.35\textwidth]{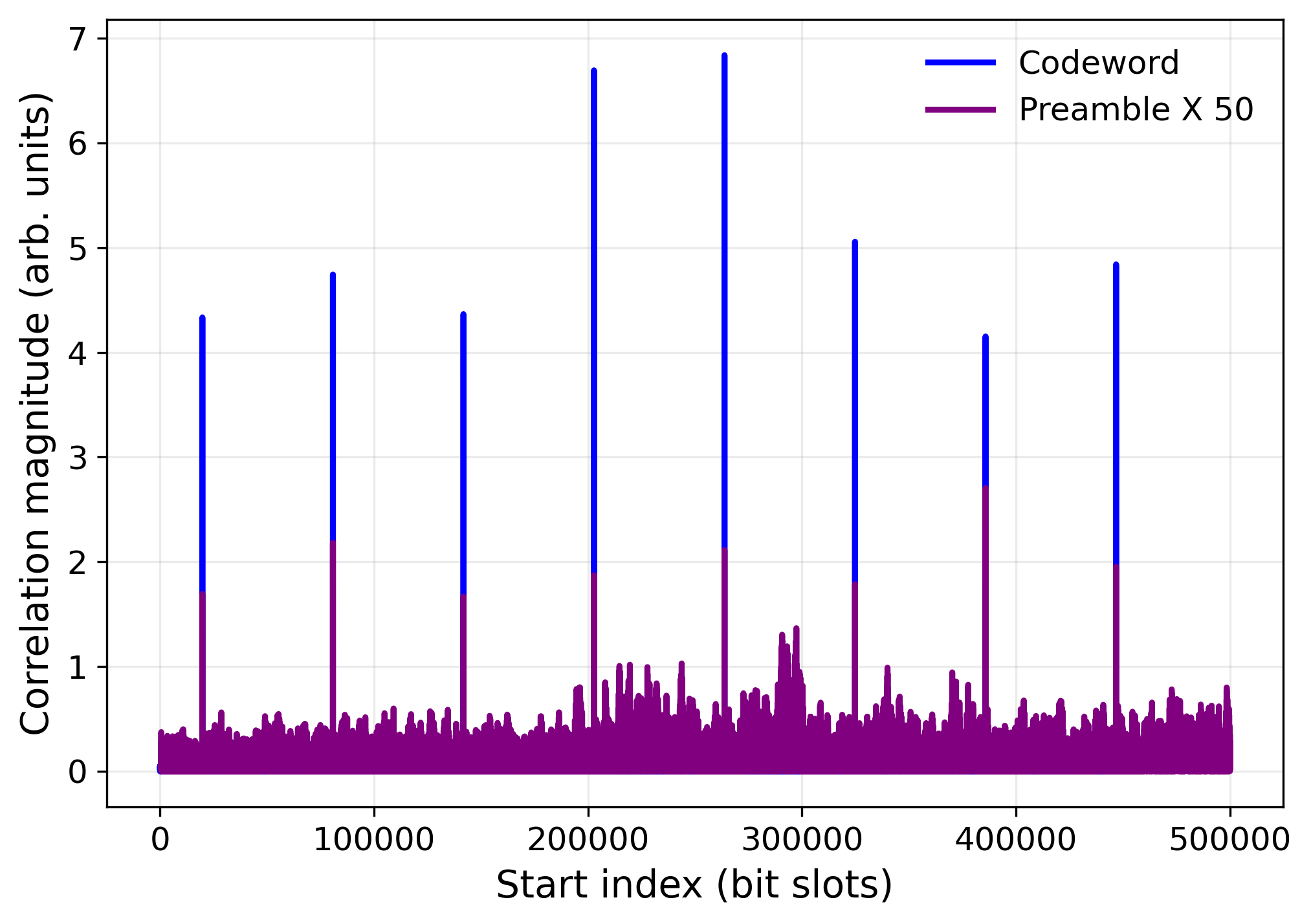}
\caption{Absolute correlation as a function of the start index for one 10 s dataset. Blue: correlation with the full codeword. Purple: correlation using only the 500 bits of the preamble (scaled for visibility). While the peak positions coincide, the preamble-based correlation exhibits significantly reduced peak amplitude and contrast relative to the noise floor, as expected.}
\label{fig:correlation}
\end{figure}

Figure~\ref{fig:correlation} shows the absolute correlation between the detected bit sequence and the reference sequence as a function of the start index for case A and B. Clear and aligned correlation peaks are observed for both methods, with the 500 bit preamble yielding accurate identification of the codeword start consistent with the full codeword case. Although the preamble signal-to-noise ratio is reduced, it remains sufficiently high to reliably detect block start times.%, while, for shorter preambles, the peaks would no longer be clearly distinguishable from the noise floor, and reliable identification of some blocks fails.

%Figure~\ref{fig:correlation} shows the absolute correlation between the detected bit sequence and the known reference sequence as a function of the start index. The clear and isolated peaks observed determine the starting index of each transmitted codeword for both methods. However, when only 500 preamble bits are used, the correlation peaks are substantially reduced in amplitude compared to the full codeword case, and their separation from the noise floor becomes significantly smaller. This result shows that the preamble alone is sufficient to identify block start times, but, as tested, for smaller preamble lengths, some blocks can no longer be reliably identified.

\begin{figure}[ht]
  \centering
  \includegraphics[width=0.4\textwidth]{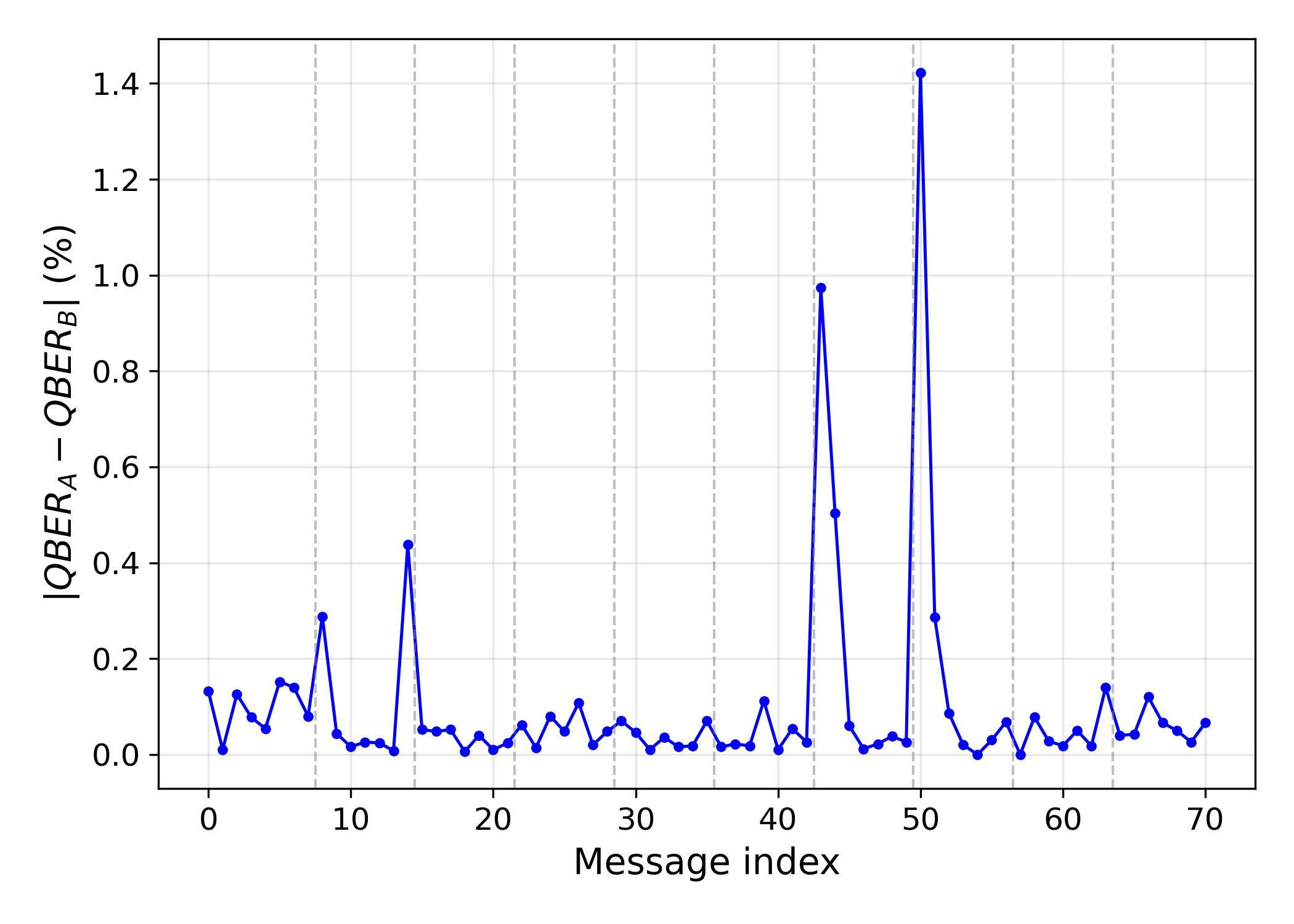} 
  \caption{Absolute difference between the QBER obtained with codeword-based synchronization ($QBER_A$) and preamble-based synchronization ($QBER_B$) for each codeword. Each point corresponds to one of the 71 blocks in the dataset. Dashed vertical lines indicate boundaries between consecutive transmission sets.}
  \label{fig:deviation}
\end{figure}

To quantify the impact of synchronization on decoding performance, we compute the QBER for both alignment strategies across all detected blocks. Bit decisions are obtained by comparing the photon counts registered in the two detectors within each bit interval, and in the case of equal counts, the bit is assigned the value 1, following the tie-breaking rule adopted in~\cite{Ploarization_QKPC}. Figure~\ref{fig:deviation} shows the absolute QBER deviation between methods A and B. The mean QBER obtained with both methods is identical within statistical uncertainty ($1.8\%$), while the average absolute deviation is only $0.1\%$, more than an order of magnitude smaller than the QBER itself. This confirms that the 500 bit preamble synchronization performs comparably to the codeword-based benchmark and contributes negligibly to the measured QBER.

\begin{figure}[ht]
  \centering
  \includegraphics[width=0.38\textwidth]{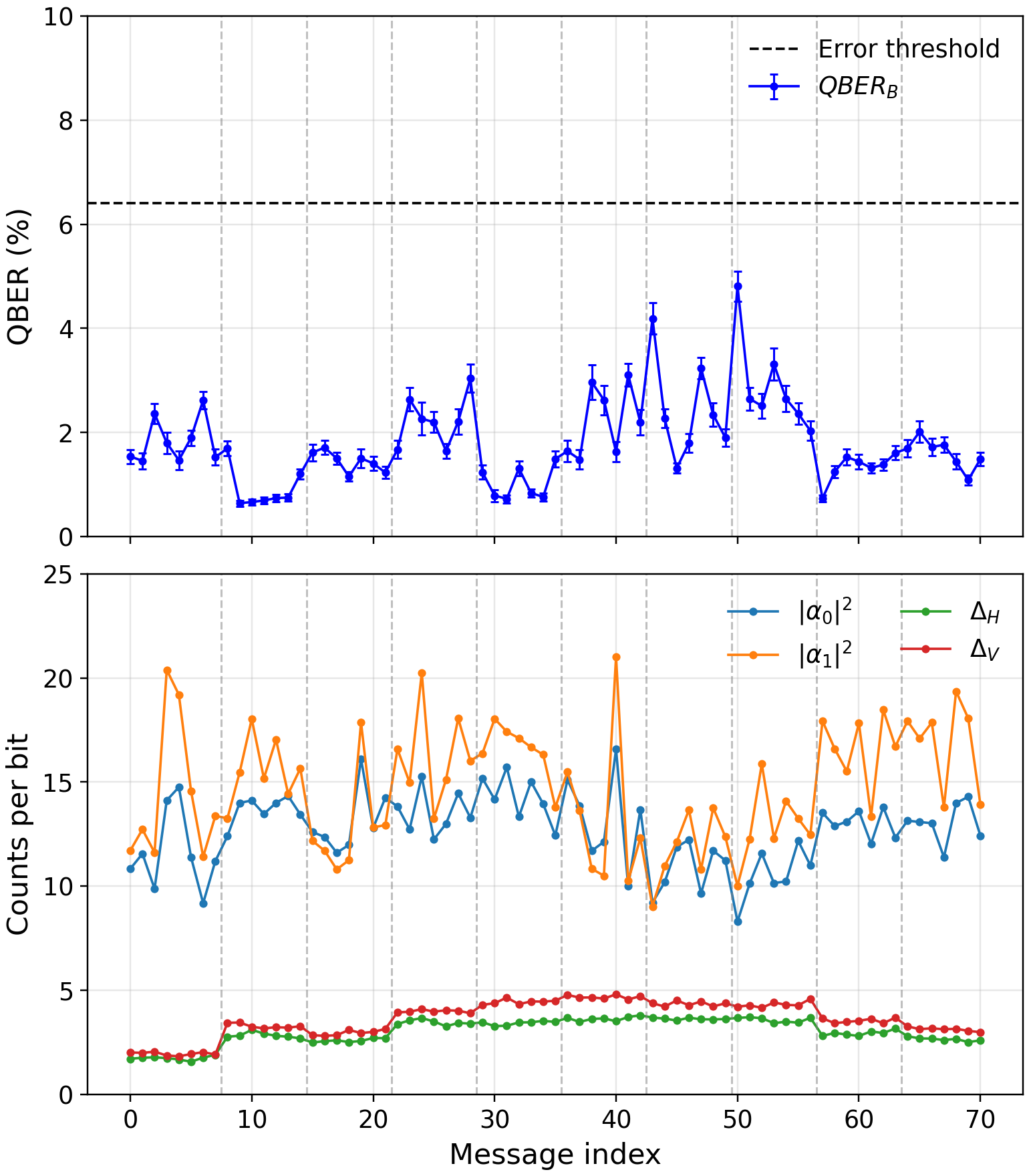} 
  \caption{Top figure: QBER obtained using preamble-based synchronization ($QBER_B$) for each of the 71 decoded blocks. Error bars represent the within-codeword standard deviation of QBER computed over 100 sub-blocks of 500 bits. Bottom figure: Estimated mean detected signal counts per bit for each state ($|\alpha|_0^2$, $|\alpha|_1^2$) together with the corresponding background levels for each detector ($\Delta_H$, $\Delta_V$) measured during idle intervals. Dashed vertical lines indicate boundaries between consecutive transmission sets.}
  \label{fig:qber}
\end{figure}

Figure~\ref{fig:qber} shows the per-codeword QBER obtained using preamble-based synchronization (top figure) together with the corresponding signal and background levels (bottom figure). For each decoded block, we estimate the mean detected signal counts per bit for each state, $|\alpha|_0^2$, $|\alpha|_1^2$, (calculated using the measured data), as well as the background contributions ($\Delta_H$, $\Delta_V$) for each channel estimated during the idle intervals between bursts.

\begin{table}[ht]
\centering
\begin{tabular}{lcc}
\toprule
Quantity & Value\\
\midrule
$|\alpha|_0^2$      & $13 \pm 2$ (counts/bit) \\
$|\alpha|_1^2$      & $15 \pm 3$ (counts/bit) \\
$\Delta_H$   & $3.0  \pm 0.6$ (counts/bit) \\
$\Delta_V$   & $3.7  \pm 0.8$ (counts/bit)\\
$\theta$    &   $59 \pm  2$ (degree)  \\
$\Omega$    &   $22 \pm  2$ (degree)  \\
\bottomrule
\end{tabular}
\caption{Experimentally determined mean signal and background counts per bit, together with effective polarization parameters, across 71 decoded blocks. Reported uncertainties correspond to block-to-block standard deviations.}
\label{tab:signal_noise_means}
\end{table}

 %For all decoded blocks, the measured QBER remains below x\% for the LDPC error correcting codes implemented, indicating that classical post-processing allows reliable recovery of the transmitted messages. 

%The experiment validates the feasibility of direct message transmission using the employed synchronization and decoding strategy.

%The experiment was designed to operate at approximately 15 detected photons per bit per channel with a polarization separation of 60°. Under these idealized conditions, the theoretical model predicts a Bob QBER of approximately 0.2\%, an Eve discrimination error of about 5.9\%, for a $\gamma=0.05$, and a maximum private capacity of roughly 0.55 bits per channel use \cite{Ploarization_QKPC}. Note that this value of $\gamma$ was chosen so as to accommodate the low coding rate of the error-correction used, further explained below.

The experimentally observed signal statistics reveal an imbalance between the two detection channels and larger-than-expected fluctuations, likely caused by misalignment during acquisition. In addition, a polarization rotation error of approximately $7^\circ$ is observed in the measurement ($\Omega_{\text{ideal}}=15^{\circ}$). Recomputing the theoretical model using the measured parameters from Table~\ref{tab:signal_noise_means} yields a predicted Bob QBER of approximately 1.1\%, an Eve discrimination error of about 15.3\% for a $\gamma=0.05$, and a maximum private capacity of roughly 0.55 bits per channel use \cite{Ploarization_QKPC}. The value of $\gamma$ was chosen to accommodate the low coding rate of the error-correction used.

For comparison, we evaluated OOK-QKPC under the same measured background-noise conditions, value of $\gamma$, and optimizing over the received photon number. The maximum private capacity remains zero, showing that OOK-QKPC would not be secure in this daylight regime.

The difference between the ideal values and the experimental parameters used can be attributed to misalignment. The transmitter was mounted on a movable support on a rooftop floor exposed to wind and mechanical vibrations, leading to beam wander and slow polarization drifts. In a practical deployment, rigid mounting, environmental shielding, and active stabilization would significantly reduce these fluctuations.

In order to enable correcting the estimated QBER, we chose a regular (50000, 25000, 5, 10) LDPC code with a rather low coding rate of $0.5$, a block length of $n=50000$, and the number of variable and check nodes as 5 and 10, respectively \cite{pyLDPC}. Using belief propagation based decoding with a maximum number of iterations set to 500 \cite{bpldpc}, we could correct up to 6.4\% of errors. As mentioned earlier, we also used a 25000 bit public random seed $s$ to construct a Toeplitz matrix, used to mask the secret message $M$ of length $k=6000$. Therefore, the wiretap coding rate, $R=k/n$, of our implementation resulted in the transmission rate of $0.12$. This rate is below the optimal rate bounded by $C_P = 0.55$ mainly due to the chosen error-correcting code, and can be greatly improved.

%However, the experimentally observed signal statistics reveal an imbalance between the two detection channels and larger-than-expected fluctuations, likely caused by misalignment during acquisition. In addition, a polarization rotation error of approximately $7^\circ$ is observed in the measurement ($\Omega_{\text{ideal}}=15^{\circ}$). Recomputing the theoretical model using the measured parameters from Table~\ref{tab:signal_noise_means} yields a predicted Bob QBER of approximately 1.1\%. This value is significantly closer to the experimentally observed performance of (1.8 $\pm$ 0.8)\%. \textcolor{red}{Add Eve error here}

% =====================================================
\section{Conclusion}
\label{sec:SectionV}
% =====================================================

We experimentally demonstrated the complete implementation of a polarization-multiplexed quantum keyless private communication protocol in a 90 meter free-space link under daylight and strong background noise conditions. The system maintained the expected performance, achieving positive private information and fully correctable and reliable message transmission. These results confirm the robustness of the approach under realistic free-space conditions and highlight the intrinsic simplicity of the protocol, showing that secure quantum communication can be achieved with reduced experimental complexity compared to QKD, even under intense ambient light.

Future work will focus on increasing the communication speed, making the system operate in real time and extending the transmission distance beyond the current demonstration. Improvements in robustness against atmospheric turbulence will be pursued through the use of active pointing and tracking systems. Further system optimization will aim at enhancing stability and scalability, paving the way toward practical long-distance free-space implementations and potential space-based deployments of QKPC.

\begin{acknowledgments}
This work is funded by national funds through FCT (Fundação para a Ciência e a Tecnologia), I.P., and, when eligible, co-funded by EU funds under project/support UID/50008/2025 - Instituto de Telecomunicações, DOI https://doi.org/10.54499/UID/50008/2025, and by the European Union through the Iberian Quantum Communication Infrastructure project, IberianQCI (CEF-DIG-2024-EUROQCI-WORKS, Grant No. 101249593).
This work was supported by the Galician Regional Government through the consolidation of Research Units: AtlantTIC, by MICIU with funding from the European Union NextGenerationEU (PRTR-C17.I1), by the Galician Regional Government with own funding through the ``Planes Complementarios de I+D+I con las Comunidades Autónomas'' in Quantum Communication, and by the European Union's Horizon Europe Framework Programme under the project ``Quantum Security Networks Partnership'' (QSNP, Grant Agreement No. 101114043).
P.N.M. acknowledges the support of FCT through scholarship 2024.01717.BD. D.R. acknowledges support from grant RYC2024-050943-I, funded by the Spanish Ministry of Science, Innovation and Universities (MICIU), the State Research Agency (AEI; 10.13039/501100011033), and the European Social Fund Plus (ESF+).
P.N.M., P.Y., and E.Z.C. would like to thank and dedicate this article to the memory of Prof. Carlos Salema, for insightful discussions, helping arrange a location for the experiment, encouragement, and for sharing his expertise and passion for error correction.
\end{acknowledgments}

\section*{Conflict of Interest}

The authors declare no competing interests.

\section*{Data Availability}

The data that support the findings of this study are available from the corresponding author upon reasonable request.

% =====================================================
\bibliographystyle{apsrev4-2}
\bibliography{qkpc_rooftop} % qkpc_rooftop.bib

@article{krause2025clock,
  title={Clock-offset recovery with sublinear complexity enables synchronization on low-level hardware for quantum key distribution},
  author={Krause, Jan and Walenta, Nino and Hilt, Jonas and Freund, Ronald},
  journal={Physical Review Applied},
  volume={23},
  number={4},
  pages={044015},
  year={2025},
  doi = {10.1103/PhysRevApplied.23.044015},
  publisher={APS}
}

@ARTICLE{QKPC_Decoy,
  author = {V\'azquez-Castro, A. and Winter, A. and Zbinden, H.},
  journal={IEEE Transactions on Information Forensics and Security}, 
  title={Quantum Keyless Private Communication With Decoy States for Space Channels}, 
  year={2024},
  volume={19},
  number={},
  pages={6213-6224},
  doi={10.1109/TIFS.2024.3410132}}

@article{UHF,
  title={Universal hashing for information-theoretic security},
  author={Tyagi, Himanshu and Vardy, Alexander},
  journal={Proceedings of the IEEE},
  volume={103},
  number={10},
  pages={1781--1795},
  year={2015},
  doi = {10.1109/JPROC.2015.2462774},
  publisher={IEEE}
}

@article{Original_QKPC_paper,
  title = {Quantum Keyless Private Communication Versus Quantum Key Distribution for Space Links},
  author = {V\'azquez-Castro, A. and Rusca, D. and Zbinden, H.},
  journal = {Phys. Rev. Appl.},
  volume = {16},
  issue = {1},
  pages = {014006},
  numpages = {12},
  year = {2021},
  month = {Jul},
  publisher = {American Physical Society},
  doi = {10.1103/PhysRevApplied.16.014006},
}

@article{Cubesat_QKPC,
  title={Optical payload design for downlink quantum key distribution and keyless communication using CubeSats},
  author={Mendes, Pedro Neto and Teixeira, Gon{\c{c}}alo Lobato and Pinho, David and Rocha, Rui and Andr{\'e}, Paulo and Niehus, Manfred and Faleiro, Ricardo and Rusca, Davide and Cruzeiro, Emmanuel Zambrini},
  journal={EPJ Quantum Technology},
  volume={11},
  number={1},
  pages={48},
  year={2024},
  doi = {10.1140/epjqt/s40507-024-00254-w},
  publisher={Springer Berlin Heidelberg}
}

@article{Ploarization_QKPC,
  title = {Quantum keyless private communication under intense background noise},
  author={Pedro Neto Mendes and Davide Rusca and Hugo Zbinden and Emmanuel Zambrini Cruzeiro},
  journal = {Phys. Rev. A},
  volume = {113},
  issue = {2},
  pages = {022624},
  numpages = {11},
  year = {2026},
  month = {Feb},
  publisher = {American Physical Society},
  doi = {10.1103/q269-nmbv},
}

@article{Qbit_4_sync,
   title={Fast and Simple Qubit-Based Synchronization for Quantum Key Distribution},
   volume={13},
   ISSN={2331-7019},
   DOI={10.1103/physrevapplied.13.054041},
   number={5},
   journal={Physical Review Applied},
   publisher={American Physical Society (APS)},
   author={Calderaro, Luca and Stanco, Andrea and Agnesi, Costantino and Avesani, Marco and Dequal, Daniele and Villoresi, Paolo and Vallone, Giuseppe},
   year={2020},
   month=may }

@article{csiszar,
  title={Broadcast channels with confidential messages},
  author={Csisz{\'a}r, Imre and Korner, Janos},
  journal={IEEE transactions on information theory},
  volume={24},
  number={3},
  pages={339--348},
  year={2003},
  doi = {10.1109/TIT.1978.1055892},
  publisher={IEEE}
}

@article{wyner,
  title={The wire-tap channel},
  author={Wyner, Aaron D},
  journal={Bell system technical journal},
  volume={54},
  number={8},
  pages={1355--1387},
  year={1975},
  doi = {10.1002/j.1538-7305.1975.tb02040.x},
  publisher={Wiley Online Library}
}

@article{Optica_free_space_QKD,
author = {Wen-Qi Cai and Yang Li and Bo Li and Ji-Gang Ren and Sheng-Kai Liao and Yuan Cao and Liang Zhang and Meng Yang and Jin-Cai Wu and Yu-Huai Li and Wei-Yue Liu and Juan Yin and Chao-Ze Wang and Wen-Bin Luo and Biao Jin and Chao-Lin Lv and Hao Li and Lixing You and Rong Shu and Ge-Sheng Pan and Qiang Zhang and Nai-Le Liu and Xiang-Bin Wang and Jian-Yu Wang and Cheng-Zhi Peng and Jian-Wei Pan},
journal = {Optica},
number = {5},
pages = {647--652},
publisher = {Optica Publishing Group},
title = {Free-space quantum key distribution during daylight and at night},
volume = {11},
month = {May},
year = {2024},
doi = {10.1364/OPTICA.511000},
}

@article{Optica_simulation_improve_qkd_satellite,
author = {Cameron Simmons and Peter Barrow and Ross Donaldson},
journal = {Optica Quantum},
number = {5},
pages = {381--389},
publisher = {Optica Publishing Group},
title = {Dawn and dusk satellite quantum key distribution using time- and phase-based encoding and polarization filtering},
volume = {2},
month = {Oct},
year = {2024},
doi = {10.1364/OPTICAQ.527880},
}

@article{Nature_daylight_long_range_QKD,
  title={Long-distance free-space quantum key distribution in daylight towards inter-satellite communication},
  author={Liao, Sheng-Kai and Yong, Hai-Lin and Liu, Chang and Shentu, Guo-Liang and Li, Dong-Dong and Lin, Jin and Dai, Hui and Zhao, Shuang-Qiang and Li, Bo and Guan, Jian-Yu and others},
  journal={Nature Photonics},
  volume={11},
  number={8},
  pages={509--513},
  year={2017},
  doi = {10.1038/nphoton.2017.116},
  publisher={Nature Publishing Group UK London}
}

@article{Nature_QKD_satellite_analysis,
  title={Finite key performance of satellite quantum key distribution under practical constraints},
  author={Sidhu, Jasminder S and Brougham, Thomas and McArthur, Duncan and Pousa, Roberto G and Oi, Daniel KL},
  journal={Communications Physics},
  volume={6},
  number={1},
  pages={210},
  year={2023},
  doi = {10.1038/s42005-023-01299-6},
  publisher={Nature Publishing Group UK London}
}

@article{science_QKD_satellite_analysis,
  title={Real-time gigahertz free-space quantum key distribution within an emulated satellite overpass},
  author={Roger, Thomas and Singh, Ravinder and Perumangatt, Chithrabhanu and Marangon, Davide G and Sanzaro, Mirko and Smith, Peter R and Woodward, Robert I and Shields, Andrew J},
  journal={Science Advances},
  volume={9},
  number={48},
  pages={eadj5873},
  year={2023},
  doi = {10.1126/sciadv.adj5873},
  publisher={American Association for the Advancement of Science}
}

@article{Nature_daylight_free_QKD,
  title={Full daylight quantum-key-distribution at 1550 nm enabled by integrated silicon photonics},
  author={Avesani, M and Calderaro, L and Schiavon, M and Stanco, A and Agnesi, C and Santamato, A and Zahidy, M and Scriminich, A and Foletto, G and Contestabile, G and others},
  journal={npj Quantum Information},
  volume={7},
  number={1},
  pages={93},
  year={2021},
  doi = {10.1038/s41534-021-00421-2},
  publisher={Nature Publishing Group UK London}
}

@article{QKD_review,
  title={The evolution of quantum key distribution networks: On the road to the qinternet},
  author={Cao, Yuan and Zhao, Yongli and Wang, Qin and Zhang, Jie and Ng, Soon Xin and Hanzo, Lajos},
  journal={IEEE Communications Surveys \& Tutorials},
  volume={24},
  number={2},
  pages={839--894},
  year={2022},
  doi = {10.1109/COMST.2022.3144219},
  publisher={IEEE}
}

@article{hayashi2011exponential,
  title={Exponential decreasing rate of leaked information in universal random privacy amplification},
  author={Hayashi, Masahito},
  journal={IEEE Transactions on Information Theory},
  volume={57},
  number={6},
  pages={3989--4001},
  year={2011},
  doi = {10.1109/TIT.2011.2110950},
  publisher={IEEE}
}

@misc{pyLDPC,
  author       = {Hicham Janati},
  title        = {pyldpc 0.7.9},
  howpublished = {\url{https://pypi.org/project/pyldpc/}},
  year         = {2020},
  note         = {Accessed: 2025-05-17}
}

@misc{bpldpc,
  author       = {Joschka Roffe},
  title        = {ldpc 2.1.0},
  howpublished = {\url{https://software.roffe.eu/ldpc/bp_decoding_example.html}},
  year         = {2023},
  note         = {Accessed: 2025-05-17}
}

\end{document}